\documentclass[twocolumn,preprintnumbers,nofootinbib,prd]{revtex4}

\usepackage{graphicx}
\usepackage{xcolor}
\usepackage[hypertex]{hyperref}
\usepackage{amsmath, amssymb}

\newcommand{\lsim}{\lesssim}
\newcommand{\ord}[1]{\mathcal{O}{(#1)}}

\newcommand{\gsim}{\gtrsim}

\newcommand{\beq}{\begin{equation}}
\newcommand{\eeq}{\end{equation}}
\newcommand{\bea}{\begin{eqnarray}}
\newcommand{\eea}{\end{eqnarray}}

\newcommand{\eps}{\varepsilon}

\newcommand{\mzd}{m_{Z_d}}
\newcommand{\sbsd}{\sin\beta \sin\beta_d}
\newcommand{\sbsdSqr}{\sin^2\beta \sin^2\beta_d}
\newcommand{\br}{{\rm BR}}
\newcommand{\tev}{{\rm TeV}}
\newcommand{\gev}{{\rm GeV}}
\newcommand{\mev}{{\rm MeV}}

\newcommand{\mat}[2][ccccc]{\left( \begin{array}{#1} #2\\ \end{array}\right)}

\begin{document}
\pagestyle{plain}
\title{``Dark'' {\boldmath $Z$} implications for Parity Violation, Rare Meson Decays, and Higgs Physics}
\author{Hooman Davoudiasl\footnote{email: hooman@bnl.gov}}
\author{Hye-Sung Lee\footnote{email: hlee@bnl.gov}}
\author{William J. Marciano\footnote{email: marciano@bnl.gov}}
\affiliation{Department of Physics, Brookhaven National Laboratory, Upton, NY 11973, USA}

\begin{abstract}
General consequences of mass mixing between the ordinary 
$Z$ boson and a relatively light $Z_d$ boson, the ``dark'' $Z$, 
arising from a $U(1)_d$ gauge symmetry, associated 
with a hidden sector such as dark matter, are examined.
New effects beyond kinetic mixing are emphasized.
$Z$-$Z_d$ mixing introduces a new source of low energy parity 
violation well explored by possible future atomic parity violation and planned polarized 
electron scattering experiments.
Rare $K (B)$ meson decays into $\pi (K) \ell^+ \ell^-$ ($\ell = e$, $\mu$) and 
$\pi (K) \nu \bar\nu$ are found to already place tight constraints on the size of $Z$-$Z_d$ mixing.
Those sensitivities can be further improved with future dedicated searches at $K$ and $B$ factories as well as binned studies of existing data.  
$Z$-$Z_d$ mixing can also lead to the Higgs decay $H \to Z Z_d$, followed by $Z \to \ell_1^+ \ell_1^-$ and $Z_d \to \ell_2^+ \ell_2^-$ or ``missing energy'', providing a potential hidden sector discovery channel at the LHC.
An illustrative realization of these effects in a 2 Higgs doublet model is presented.
\end{abstract}
\maketitle

\section{Introduction}
The existence of cosmic dark matter is now essentially established.
It appears to constitute about 22\% of the energy-matter budget of the Universe, significantly more than the 4\% attributed to visible matter \cite{PDG}.
Nevertheless, the exact nature of dark matter remains mysterious.
Is it mainly a new, cosmologically stable, elementary particle that interacts with our visible world primarily through gravity or does it have weak interaction properties that allow it to be detected at high energy accelerators or in sensitive underground cryogenic experiments?
Both avenues of exploration are currently in progress.
A discovery would revolutionize our view of the Universe and the field of elementary particle physics.

Recently, a possible generic new property of dark matter has been 
postulated \cite{DarkMatter} to help explain various astrophysical 
observations of positron excesses \cite{Adriani:2008zr}.  
The basic idea is to introduce a new $U(1)_d$ gauge symmetry 
mediated by a relatively light $Z_d$ boson that couples to the ``dark'' charge of hidden sector states, 
an example of which is dark matter.
Such a boson has been dubbed the ``dark'' photon, secluded or hidden boson, {\it etc} \cite{EarlyStudy}.
Within the framework adopted in our work, however, we refer to it as the ``dark'' $Z$ 
because of its close relationship to the ordinary $Z$ of the Standard Model (SM) via $Z$-$Z_d$ mixing.
Consequences of that mixing will be explored in this paper, where after describing the basic characteristics of the dark $Z$, we provide constraints on its properties imposed by low energy parity violating experiments such as atomic parity violation and polarized electron scattering.
Future sensitivities are also discussed.
We then briefly describe bounds on the mixing currently obtained from rare $K$ and $B$ decays along with the potential for future improvements.

Perhaps the most novel prediction from $Z$-$Z_d$ mixing is its implications for high energy experiments.
In particular, it leads to a potentially observable new type of Higgs decay, $H \to Z Z_d$, with pronounced discovery signatures that we describe \cite{heavyZprime}.
We also discuss a 2 Higgs doublet (2HD) model that exhibits all the features of our general $Z$-$Z_d$ mixing scenario.
(Some works of similar spirit, but different contexts can be found in, for example, Refs.~\cite{Babu:1997st,Uboson,Frandsen:2011cg,Lebedev:2011iq,Kamenik:2011vy}.)

\section{Set Up}
We begin with what might be called the usual ``dark'' boson scenario.
It is assumed that a new $U(1)_d$ gauge symmetry of the dark matter or any hidden sector interacts with the 
$SU(3)_C \times SU(2)_L \times U(1)_Y$ of the SM via kinetic mixing between $U(1)_Y$ and $U(1)_d$ \cite{Holdom:1985ag}.
That effect is parametrized by a gauge invariant $B_{\mu\nu}Z_d^{\mu\nu}$ interaction
\beq
\begin{split}
{\cal L}_\text{gauge} &= -\frac{1}{4} B_{\mu\nu} B^{\mu\nu} + \frac{1}{2} \frac{\eps}{\cos\theta_W} B_{\mu\nu} Z_d^{\mu\nu} - \frac{1}{4} Z_{d \mu\nu} Z_d^{\mu\nu} \\
B_{\mu\nu} &= \partial_\mu B_\nu - \partial_\nu B_\mu \qquad Z_{d \mu\nu} = \partial_\mu {Z_d}_\nu - \partial_\nu {Z_d}_\mu
\end{split}
\eeq
with $\eps$ a dimensionless parameter that is unspecified (the normalization of the term proportional 
to $\eps$ has been chosen to simplify the notation in the results that follow). 
At the level of our discussion, $\eps$ is a potentially infinite counter term necessary for renormalization.
Its finite renormalized value is to be determined by experiment.
In most discussions, $\eps$ is assumed to be $\lsim {\cal O} ({\rm few}\times10^{-3})$.
It could, of course, be much smaller \cite{Abel:2008ai}.

After removal of the $\eps$ cross-term by field redefinitions 
\beq
B_\mu     \to B_\mu + \frac{\eps}{\cos\theta_W} {Z_d}_\mu
\label{eq:2}
\eeq
leading to
\beq
\begin{split}
A_\mu &\to A_\mu + \eps {Z_d}_\mu \\
Z_\mu &\to Z_\mu - \eps \tan\theta_W {Z_d}_\mu
\end{split}
\label{eq:2a}
\eeq
for the photon and $Z$ boson fields, one is left with an induced coupling of the $Z_d$ to the usual electromagnetic current (with summation over all charged quarks and leptons)
\beq
\begin{split}
{\cal L}_\text{int} &= - e \eps J_{em}^\mu {Z_d}_\mu \\
         J_{em}^\mu &= \sum_f Q_f \bar f \gamma^\mu f + \cdots
\end{split}
\label{eq:3}
\eeq
where the ellipsis includes $W^\pm$ current terms and $Q_f$ is the electric charge ($Q_e = -1$).
(It is generally assumed that $U(1)_d$ is broken and $Z_d$ becomes massive via a scalar Higgs singlet or a Stueckelberg mass generating mechanism \cite{Stueckelberg:1900zz,Feldman:2007wj}.)
Note also that the induced coupling of $Z_d$ to the weak neutral current 
via Eq.~(\ref{eq:2a}) is highly suppressed at low energies in the above basic scenario because of a cancellation between $\eps$ dependent field redefinition and $Z$-$Z_d$ mass matrix diagonalization effects induced by $\eps$ (see, for example, 
Ref.~\cite{Gopalakrishna:2008dv} and our Appendices A and B).

The phenomenology of the interaction in Eq.~\eqref{eq:3} has been well examined as a function of $m_{Z_d}$ and $\eps$ ({\it e.g.} Refs.~\cite{Batell:2009yf,Bjorken:2009mm,Jaeckel:2010ni}).
With the assumption $10 ~\mev \lsim m_{Z_d} \lsim 10 ~\gev$ and $\eps \lsim {\cal O}(\text{few} \times 10^{-3})$, bounds have been given and new experiments are underway to find the $Z_d$ via its production in high intensity electron scattering \cite{Abrahamyan:2011gv}.  
We will consider this same mass range for our phenomenological analysis in this work.
The lower bound $m_{Z_d} \gsim 10 ~\mev$ is required in order that astrophysical and beam-dump processes do not severely constrain the interactions of dark $Z$ which, as discussed below, develops an axionlike component for $m_{Z_d} \to 0$.

Because of its coupling to our particle world via the small electromagnetic current coupling in Eq.~\eqref{eq:3}, $Z_d$ is often called the ``dark'' photon (even though that name was originally intended for a new weakly coupled long-range interaction \cite{Ackerman:2008gi}).

Here, we generalize the above $U(1)_d$ kinetic mixing scenario to include $Z$-$Z_d$ mass 
mixing by introducing the $2 \times 2$ mass matrix
\beq
M_0^2 = m_Z^2 \begin{pmatrix}1 & -\eps_Z \\ 
-\eps_Z  & ~~m_{Z_d}^2/m_Z^2\end{pmatrix}
\label{eq:new4}
\eeq
where $m_{Z_d}$ and $m_Z$ (with $m_{Z_d}^2 \ll m_Z^2$) represent the ``dark'' $Z$ and SM $Z$ masses in the limit of no mixing.
The $Z$-$Z_d$ mixing is parametrized by
\beq
\eps_Z = \frac{m_{Z_d}}{m_Z} \delta \, ,
\label{eq:6}
\eeq
with $\delta$ a small model dependent quantity.
We ignore the $\eps$ contribution from Eq.~(\ref{eq:2}) 
in the mass matrix, since its inclusion would affect this part of our discussion only at ${\cal O} (\eps^2)$ (see Appendix B).
The assumed off-diagonal $m_{Z_d}$ dependence in Eq.~(\ref{eq:6}) allows smooth $m_{Z_d} \to 0$ 
behavior for all $\eps_Z$-induced amplitudes involving $Z_d$, even those stemming from nonconserved current interactions.
Also, for simplicity, ordinary fermions are assumed to be neutral under $U(1)_d$, {\it i.e.} they do not carry any fundamental dark charge.
Their only couplings to $Z_d$ are induced through $\eps$ and $\eps_Z$.
More general cases are possible and interesting, but beyond the scope of this paper.

So far, $\delta$ is rather arbitrary, although $0 \le \delta^2 < 1$ is required to avoid an infinite-range or tachyonic $Z_d$.  
One expects $\delta$ to be small because of the disparity of $m_Z$ and $\mzd$.  
We later show that low energy phenomenology actually requires $\delta^2 \lsim 0.006$, 
while rare $K$ and $B$ decays have sensitivity to $\delta^2 \lsim 10^{-4} - 10^{-6}$ for low mass $Z_d$.  
We will also demonstrate how the form in Eq.~\eqref{eq:new4} naturally 
emerges in a simple 2HD extension of the SM, the details of which will be discussed in Appendix B.
However, we emphasize that our general 
results follow from $Z$-$Z_d$ mixing through a 
generic mass matrix of the form in Eq.~\eqref{eq:new4} 
and are not exclusively tied to any specific expanded Higgs sector.
That mixing could, for example, potentially arise from loop effects or dynamical symmetry breaking.

Overall, mixing leads to mass eigenstates $Z$ and $Z_d$
\beq
\begin{split}
Z   &= Z^0 \cos\xi - Z_d^0 \sin\xi \\
Z_d &= Z^0 \sin\xi + Z_d^0 \cos\xi
\end{split}
\eeq
where (see Appendix B)
\beq
\begin{split}
\tan 2\xi  &\simeq 2\frac{m_{Z_d}}{m_Z} \delta = 2 \eps_Z \, .
\end{split}
\eeq
It is expected that $\sin\xi$ is very small (partly because of the assumed smallness of $m_{Z_d} / m_Z$ and partly because of small $\delta$) and does not measurably affect $Z$ pole parameters (such as $m_Z$ and $\Gamma_Z$) because these are shifted fractionally at $\ord{\eps_Z^2}$, and require only $\eps_Z\lsim \ord{0.01}$.  
However, it can, nevertheless, lead to other interesting new 
phenomenology which overcomes the $m_{Z_d} / m_Z$ suppression 
in $\eps_Z$.

As the first example, we consider very low $Q^2$ parity violating effects where the smallness of $m_{Z_d} / m_Z$ in the induced $Z_d$ couplings is offset by the $m_Z^2 / m_{Z_d}^2$ enhancement from $Z$ vs $Z_d$ propagators.
Then we describe the induced decays $K\to \pi Z_d$ and $B\to K Z_d$, as well as the 
high energy decay $H \to Z Z_d$, where the small induced coupling factor $m_{Z_d} / m_Z$ is overcome by $m_K / m_{Z_d}$, $m_B / m_{Z_d}$ and $m_H / m_{Z_d}$ enhancements, respectively, in the longitudinal polarization component of the $Z_d$ production amplitudes.

\section{Atomic Parity Violation and Polarized Electron Scattering}
We begin our analysis by writing out the full $Z_d$ coupling to fermions from $\eps$ as well as $\eps_Z$.
\beq
{\cal L}_\text{int} = \left( -e \eps J_\mu^{em} - \frac{g}{2 \cos\theta_W} \eps_Z J_\mu^{NC} \right) Z_d^\mu
\label{eq:7}
\eeq
where $J_\mu^{em}$ is given in Eq.~\eqref{eq:3} and
\beq
J_\mu^{NC} = \sum_f ( T_{3f} - 2 Q_f \sin^2\theta_W ) \bar f \gamma_\mu f - T_{3f} \bar f \gamma_\mu \gamma_5 f
\label{eq:8}
\eeq
with $T_{3f} = \pm 1/2$ ($T_{3e} = - 1/2$) and $\sin^2\theta_W \simeq 0.23$ is the weak mixing angle of the SM.
The inclusion of $Z$-$Z_d$ mixing has introduced parity violation.
The $J_\mu^{NC} Z_d^\mu$ coupling is similar to the $J_\mu^{NC} Z^\mu$ coupling of the SM $Z$ but reduced by $\eps_Z$ in magnitude.
Hence, the name ``dark'' $Z$, since it is the $\eps_Z$ induced interactions that we primarily address.
Note that the effects of $\eps$ and $\eps_Z$ can be combined into a simple form
\beq
{\cal L}_\text{int} = - \frac{g}{2 \cos\theta_W} \eps_Z J_\mu^{NC '} Z_d^\mu
\label{eq:9}
\eeq
by the replacement $J_\mu^{NC '} (\sin^2\theta_W) = J_\mu^{NC} (\sin^2 \theta'_W)$
\beq
\sin^2\theta'_W = \sin^2 \theta_W - \frac{\eps}{\eps_Z} \cos\theta_W \sin\theta_W
\label{eq:10}
\eeq
in Eq.~\eqref{eq:8}.
In that format, one can judge the relative importance of $\eps$ in low energy $Z_d$ phenomenology.
It depends on the size of $(\eps / \eps_Z) (\cos\theta_W / \sin\theta_W)$.
For $\eps$ very small, it has little effect, but will be significant if $\eps \sim \eps_Z$.

The new source of parity violation in Eq.~\eqref{eq:7} or Eq.~\eqref{eq:9}, 
is particularly important for experiments at $Q^2 < m_{Z_d}^2$ where the $Z_d$ propagator can 
provide an enhancement owing to $m_{Z_d}^2 \ll m_Z^2$.
The overall effect for parity violating amplitudes ${\cal M}_{NC}^\text{PV} = (G_F / 2\sqrt{2}) F(\sin^2\theta_W)$ in the SM is (in leading order) to replace
\beq
\begin{split}
G_F &\to \rho_d G_F \\
\sin^2\theta_W &\to \kappa_d \sin^2\theta_W
\end{split}
\eeq
with \cite{Marciano:1980pb}
\beq
\begin{split}
\rho_d &= 1 + \delta^2 \frac{m_{Z_d}^2}{Q^2 + m_{Z_d}^2} \\
\kappa_d &= 1 - \frac{\eps}{\eps_Z} \delta^2 \frac{\cos\theta_W}{\sin\theta_W} \frac{m_{Z_d}^2}{Q^2 + m_{Z_d}^2}
\end{split}
\label{eq:12}
\eeq
or from Eq.~\eqref{eq:6}
\beq
\kappa_d = 1 - \eps \frac{m_Z}{m_{Z_d}} \delta \frac{\cos\theta_W}{\sin\theta_W} \frac{m_{Z_d}^2}{Q^2 + m_{Z_d}^2} \, .
\eeq

It is quite plausible that in a more complete theory, $\eps \propto (m_{Z_d} / m_Z) \delta = \eps_Z$.
Then, the effects from kinetic mixing and $Z$-$Z_d$ mixing become similar in form and magnitude.
Here, we allow $\eps$ to remain a separate independent parameter.

Assuming no accidental cancellation between the $\rho_d$ and $\kappa_d$ in Eq.~\eqref{eq:12},
Cesium atomic parity violation currently provides the best low energy experimental constraint on those parameters over the entire approximate range of interest ($10 ~\mev \lsim m_{Z_d} \lsim 10 ~\gev$) since $Q^2 \ll \mzd^2$.
The nuclear weak charge measured in atomic parity violation (to lowest order in the SM) is given by $Q_W = - N + Z (1 - 4 \sin^2 \theta_W)$ which when compared with experiment probes new physics.
There is excellent agreement between the SM prediction for the weak charge 
of Cesium (including electroweak radiative corrections) \cite{Marciano:1982mm,Marciano:1990dp,Marciano:1993jd}
\beq
Q_W^\text{SM} (^{133}_{55}\text{Cs}) = -73.16(5)
\eeq
and the experimental value \cite{Porsev:2009pr,Wood:1997zq,Bennett:1999pd}
\beq
Q_W^\text{exp} (^{133}_{55}\text{Cs}) = -73.16(35).  
\eeq
Based on the shift due to $\eps$, $\eps_Z$ and $\delta$
\beq
Q_W^\text{SM} \to -73.16 (1 + \delta^2) + 220 \frac{\eps}{\eps_Z} \delta^2 \cos\theta_W \sin\theta_W,  \nonumber
\eeq
the above agreement then implies the following constraints 
\bea
&& \left|\delta^2 (1-1.27 \frac{\eps}{\eps_Z}) \right| \lsim 0.005 ~(1 \sigma) \label{eq:17}\\
&& \delta^2 \lsim 0.006 ~(\text{one-sided}~ 90\% ~\text{C.L.}),\; {\rm for} \; \eps\ll \eps_Z. \label{eq:16} 
\eea
For $\eps \simeq \eps_Z$, the constraints on $\delta^2$ become diluted 
and the possibility of cancellation occurs if one tunes $\eps / \eps_Z \simeq 0.8$.
(We note that the fine tuning $\eps/\eps_Z \simeq 0.8$ is similar to a relation employed in Ref.~\cite{Frandsen:2011cg} to try and reconcile what appears to be discrepancies in dark matter search scattering experiments on heavy nuclei.
However, such a scenario is significantly constrained by the bounds on $\delta$ described below.)

An independent constraint primarily applicable to $\kappa_d$ because of its relative insensitivity to $\rho_d$ comes from parity violating polarized electron-electron Moller scattering asymmetries \cite{Derman:1979zc,Czarnecki:1995fw}.  
Experiment E158 at SLAC \cite{Anthony:2005pm} measured the low energy value of $\sin^2\theta_W (Q^2)$ at $Q^2 \simeq (0.16 ~\gev)^2$ and compared it with expectations based on running the $Z$ pole value $\sin^2\theta_W (m_Z)$ down to low $Q^2$ \cite{Czarnecki:1995fw}.
The good agreement with SM loop effects leads to (ignoring the small $\rho_d$ effect)
\beq
\left| \frac{\eps}{\eps_Z} \delta^2 \right| \frac{m_{Z_d}^2}{(0.16 ~\gev)^2 + m_{Z_d}^2} \lsim 0.006 \, .
\label{eq:18}
\eeq
For $m_{Z_d}^2 \gg (0.16 ~\gev)^2$ and $\eps_Z \simeq \eps$, 
the constraints in Eqs.~\eqref{eq:16} and \eqref{eq:18} are essentially the same.
However, for a light $m_{Z_d} \lsim 200 ~\mev$, the bound in Eq.~\eqref{eq:18} can be somewhat diluted.
Nevertheless, for some range of $(\eps, m_{Z_d}$) values, Eq.~\eqref{eq:18} can provide more restrictive bounds on $\delta$.
For example, consider $\eps \simeq 2 \times 10^{-3}$ and $m_{Z_d} \simeq 100 ~\mev$ which lie in the region favored by the current discrepancy between theory and experimental values of the muon anomalous magnetic moment \cite{Fayet:2007ua}.
In that case, Eq.~\eqref{eq:18} becomes
\beq
|\delta| < 0.01
\label{eedelta}
\eeq
which is considerably tighter than Eq.~\eqref{eq:16}.
If the muon anomaly discrepancy is because of a light $Z_d$ and $\eps \sim 10^{-3}$, that boson's 
effect on the value of $\sin^2\theta_W$ extracted from future more precise very low $Q^2$ parity violating experiments \cite{McKeown:2011yj} could eventually become observable.

The sensitivity in Eqs.~\eqref{eq:18} and \eqref{eedelta} 
is expected to improve by up to an order of magnitude from ongoing and proposed polarized $ep$ and $ee$ scattering experiments at JLAB \cite{McKeown:2011yj} as well as proposed $Q^2 \simeq (0.05 ~\gev)^2$ $ep$ studies at MESA in Mainz \cite{MESA}.  Our analysis illustrates the complementarity of direct searches at intense electron scattering facilities in JLAB and Mainz for a light vector particle 
(the ``dark" photon coupled through kinetic mixing) produced via electron scattering, with low $Q^2$ 
measurements of $\sin^2\theta_W$ in parity violating experiments (that probe $\eps$ and the mass mixing of the ``dark" $Z$).  
We also note that proposed measurements of atomic parity violation for ratios of different nuclear isotopes would eliminate atomic physics uncertainties as well as any dependence on $\rho_d$ \cite{Dzuba:1985gw,Fortson:1990zz,Monroe:1990zz,Brown:2008ib}.
They would then be sensitive to $(\eps / \eps_Z) \delta^2$ but with negligible $Q^2$ dependence (since $Q^2 \simeq 0$).
It is amusing to note that in principle, very low energy measurements of $\sin^2\theta_W$ in atomic parity violation and low $Q^2$ polarized electron scattering experiments could find different $\sin^2\theta_W$ results from one another if a very low mass $Z_d$ is contributing to both, because of the $Q^2$ dependence in Eq.~\eqref{eq:12}.

Our conclusion, based on the above discussion, is that currently, $\delta^2 \lsim 0.006$ 
is a modest, reasonably reliable constraint for most values of $\mzd$, 
although fine tuning of $\eps$ and $\eps_Z$ could loosen the bound.
That constraint can be much stronger for $\eps \sim 10^{-3}$ [see Eq.~\eqref{eedelta}], 
and could be further improved significantly by future low energy parity violating experiments.
For now, the bound $\delta^2 \lsim 0.006$ provides a starting 
point for comparison with the sensitivity to $\delta^2$ in rare 
$K$ and $B$ decays which we next describe.

\section{Rare {\boldmath $K$ and $B$} Decays}
\begin{figure}[t]
\begin{center}
\includegraphics[width=0.45\textwidth]{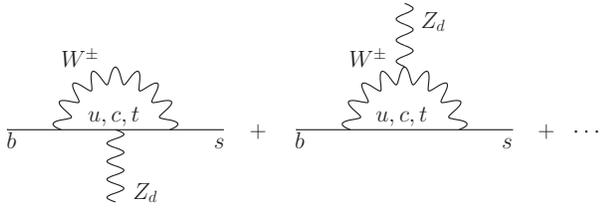}
\end{center}
\caption{Examples of diagrams contributing to $b \to s Z_d$. Similar diagrams give rise to $s \to d Z_d$.}
\label{fig:1}
\end{figure}

Experimental studies of rare flavor-changing weak neutral current decays of $K$ and $B$ mesons have proven to be powerful probes of high and low scale ``new'' physics phenomena.
Here, we illustrate the effect of $Z$-$Z_d$ mass mixing on the transition amplitudes $s \to d Z_d$ and $b \to s Z_d$ induced within the framework of Cabibbo-Kobayashi-Maskawa (CKM) charged current mixing (See Fig.~\ref{fig:1}).
Those loop induced couplings can lead to decays such as $K \to \pi Z_d$ and $B \to K Z_d$ or $K^* Z_d$ characterized by the signature $Z_d \to \ell^+ \ell^-$ ($\ell = e$ or $\mu$) with invariant mass $m_{\ell\ell} = m_{Z_d}$ or $Z_d \to$ missing energy where $Z_d$ decays into $\nu\bar\nu$ or essentially undetectable light hidden sector particles.
In all such 2-body decays, the mono energetic outgoing $\pi$ or $K$ will provide a tight constraint (for a given $m_{Z_d}$) and a very distinct overall signal.

\begin{figure}[t]
\begin{center}
\includegraphics[width=0.45\textwidth]{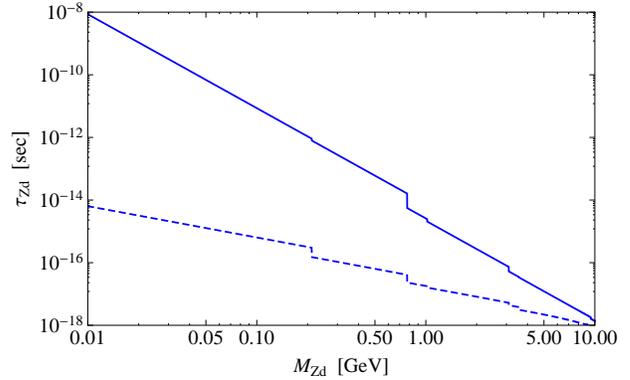}
\end{center}
\caption{$Z_d$ lifetime with $Z_d$ mass for $\delta^2 = 10^{-4}$ with $\eps = 0$ (solid blue curve) and $\eps = 2 \times 10^{-3}$ (dashed blue curve) cases. We take $\rho$, $\phi$, $J/\psi$, $\Upsilon$ masses as the representative threshold for decays to mesons.}
\label{fig:lifetime}
\end{figure}

Here, we note that the phenomenology of $Z_d$ is affected by its lifetime $\tau_{Z_d}$.
A sufficiently large value of $\tau_{Z_d}$ will allow $Z_d$ to escape the detector and lead to a missing energy signal.
However, for smaller values of $\tau_{Z_d}$, a displaced vertex can provide a distinct signature.
In Fig.~\ref{fig:lifetime}, using representative values of $\delta$ and $\eps$, we have plotted $\tau_{Z_d}$ for $10 ~\mev \le m_{Z_d} \le 10 ~\gev$, assuming that $Z_d$ only decays into SM final states.
We provide a simple formula for the partial width of $Z_d$ into SM fermions, $\Gamma(Z_d \to f \bar f)$, in Appendix C.

Of course, the amplitudes for $\bar d s Z_d$ and $\bar s b Z_d$ being loop induced will in general depend on the details of the complete model considered, including its underlying Higgs flavor symmetry breaking structure.  Those details are beyond the scope of this paper where we are primarily interested in the generic effects of $Z$-$Z_d$ mixing parametrized by $\eps_Z = (m_{Z_d}/m_Z) \delta$ in Eq.~\eqref{eq:6}.

A simple illustrative example of a scenario that leads to $Z$-$Z_d$ mixing and CKM induced flavor-changing weak neutral currents is the Type-I 2HD model discussed in Sec.~\ref{2HD} and detailed in Appendix B.
There, the underlying $U(1)_d$ gauge symmetry naturally 
forbids tree level flavor-changing neutral currents in the scalar and pseudoscalar Higgs sectors.
It also yields, through Higgs doublet and singlet vacuum expectation values, a mechanism 
to provide mass for $Z_d$ and give rise to a small $\delta$ in Eq.~\eqref{eq:6}.

To obtain the induced $Z_d$ flavor-changing amplitudes, 
we can make use of existing CKM loop induced calculations for 
$\bar d_L \gamma_\mu s_L Z^\mu$ and $\bar s_L \gamma_\mu b_L Z^\mu$ 
amplitudes \cite{dsZ} and replace $Z \to \eps_Z Z_d$. (See Fig.~\ref{fig:1}.)
(We ignore kinetic mixing induced couplings, since their effects are highly suppressed.
For example, Ref.~\cite{Batell:2009jf} found $\br (B \to K Z_d) \sim 6 \times 10^{-7} \eps^2$ for $m_{Z_d} \simeq 1 ~\gev$.
As we demonstrate, mass mixing, $\eps_Z$, induced rates can be much larger and potentially observable.)
As an alternative computational strategy, if we are primarily interested in relatively light $Z_d$ bosons compared to $m_K$ and $m_B$, we can employ the Goldstone boson equivalence theorem \cite{GET} to obtain amplitudes for longitudinally polarized $Z_d$ bosons from flavor-changing axionlike pseudoscalar couplings well documented in the literature.
For our purpose, the latter approach will suffice; however, the direct $Z$ calculation provides a nice cross-check.  
Nevertheless, we note that the results discussed below should be viewed as somewhat incomplete 
and should be taken as approximate.

The relevant $\bar d_L \gamma_\mu s_L \partial^\mu a$ and $\bar s_L \gamma_\mu b_L \partial^\mu a$ axion couplings were computed for the 2HD model more than 30 years ago by Hall and Wise \cite{Hall:1981bc} 
and independently by Frere, Vermaseren and Gavela \cite{Frere:1981cc}.
More recently, they were checked and applied to the decay 
$B \to K a$, $a \to \ell^+ \ell^-$ in Ref.~\cite{Freytsis:2009ct}.
Here, we use those results to estimate the branching ratios for $K \to \pi Z_d$ (longitudinal) and $B \to K Z_d$ (longitudinal) which should approximate the full $Z_d$ final state rates up to corrections of 
${\cal O} (m_{Z_d}^2 / m_K^2)$ and ${\cal O} (m_{Z_d}^2 / m_B^2)$ respectively.
Comparison of those estimates with experiments can then be used to constrain 
$\delta$ for the ranges $m_{Z_d}^2 \ll (m_K - m_\pi)^2$ and 
$m_{Z_d}^2 \ll (m_B - m_K)^2$ modulo regions not covered because of experimental acceptance cuts on the data (which are beyond the scope of this paper).
For example, $m_{Z_d} < 140 ~\mev$ is not covered because of $\pi^0 \to e^+ e^- \gamma$ Dalitz decay background.
Similarly, masses of $Z_d$ near charmonium resonance regions are not covered.

We begin with the predicted branching ratio for $K \to \pi Z_d$ (longitudinal) in the 2HD model.
Based on the analysis in Ref.~\cite{Hall:1981bc}, but adjusting for a modern $m_t$ value, since top now dominates the amplitudes in Fig.~\ref{fig:1}
\beq
\br (K^+ \to \pi^+ Z_d)_\text{long} \simeq 4 \times 10^{-4} \delta^2 ,
\label{eq:n1}
\eeq
where the numerical factor in that expression includes QCD 
suppression effects and depends on the physical charged scalar Higgs mass of the 2HD model.
Those uncertainties should be considered part of the overall model dependence of our analysis.

The $Z_d$ produced in Eq.~\eqref{eq:n1} is expected to decay promptly (see, however, Fig.~\ref{fig:lifetime}) to $\ell^+ \ell^-$ pairs with invariant mass $m_{Z_d}$ or to missing energy that might be $\nu \bar\nu$ or light hidden sector particles.
Those decays would add to the SM predictions and should be part of the experimentally measured branching ratios \cite{PDG,Appel:1999yq,Batley:2009aa}
\bea
&&\br (K^+ \to \pi^+ e^+ e^-)_\text{exp}     = (3.00 \pm 0.09) \times 10^{-7} \label{eq:n2a}\\
&&\br (K^+ \to \pi^+ \mu^+ \mu^-)_\text{exp} = (9.4  \pm 0.6)  \times 10^{-8} \label{eq:n2b}\\
&&\br (K^+ \to \pi^+ \nu \bar\nu)_\text{exp} = (1.7  \pm 1.1)  \times 10^{-10}\label{eq:n2c}
\eea
unless eliminated by acceptance cuts which would negate bounds in certain $m_{Z_d}$ regions.
For example, the result in Eq.~\eqref{eq:n2a} applied a $m_{ee} > 140 ~\mev$ cut 
while Eq.~\eqref{eq:n2c} was obtained with a rather stringent cut on $E_\pi$.
Clearly, a new round of bump hunting in the $\ell^+ \ell^-$ spectrum is warranted.
Toward that end, we note that $Z_d \to \ell^+ \ell^-$ decays will have a characteristic 
polarized spin-1 $\sin^2\theta$ distribution relative to the longitudinal polarization of the $Z_d$.
Unlike the spin-0 axion case, where because of chiral conservation the $a$ preferentially decays to the heaviest fermion possible and the distribution is isotropic, we expect $\br (Z_d \to e^+ e^-) \simeq \br (Z_d \to \mu^+ \mu^-)$ modulo phase space.  

With the above caveats, we compare Eq.~\eqref{eq:n1} with \eqref{eq:n2a}, \eqref{eq:n2b}, and \eqref{eq:n2c} which agree with SM expectations and find rather tight bounds
\bea
&&|\delta| \lsim 0.01 / \sqrt{\br (Z_d \to e^+ e^-)} \label{eq:n3a}\\
&&|\delta| \lsim 0.001 / \sqrt{\br (Z_d \to {\rm missing\; energy})}\label{eq:n3b}
\eea
modulo acceptance cut criteria.

Eqs.~\eqref{eq:9} and \eqref{eq:10} yield \cite{Albert:1979ix}
\beq
\frac{\br (Z_d \to e^+ e^-)}{\br (Z_d \to \nu \bar\nu)} \simeq \frac{1}{6} + 
\frac{1}{2} \left(\frac{\eps}{\eps_Z}\right)^2\,,
\label{ZdBR}
\eeq
where $\eps$ from kinetic mixing now comes into play.
For $\eps \gg \eps_Z$, the charged lepton decays dominate and Eq.~\eqref{eq:n3a} is more applicable.
For $\eps \lsim \eps_Z$, the tighter constraint in Eq.~\eqref{eq:n3b} takes precedence.
Of course, both should be used cautiously, given their model and experimental acceptance dependence.

For the case of $B \to K Z_d$ (longitudinal), we can apply a similar approach and find \cite{Hall:1981bc,Frere:1981cc,Freytsis:2009ct}
\beq
\br (B \to K Z_d)_\text{long} \simeq 0.1 \delta^2 .
\label{eq:BtoKZd}
\eeq
The relatively large coefficient in Eq.~\eqref{eq:BtoKZd} results from a factor of $m_t^4$ in the $b \to s Z_d$ loop induced correction from Fig.~\ref{fig:1}.
That factor makes rare $B$ decays a particularly sensitive probe of the $Z_d$.
Employing the recent bounds that follow from the discussion of 
$B \to K a$, with the axion-type particle  
$a \to \ell^+ \ell^-$ in Refs.~\cite{Batell:2009jf,Freytsis:2009ct} implies conservatively $\br(B \to K Z_d \to K \ell^+ \ell^-) <10^{-7}$, while
the bound from $B$-decay containing missing energy are based on \cite{PDG,Aubert:2008ps,Wei:2009zv}
\beq
\br (B^+ \to K^+ \bar\nu \nu)_\text{exp} < 1.4 \times 10^{-5} .
\label{eq:BtoKnunu}
\eeq
We then roughly find
\bea
&&|\delta| \lsim 0.001 / \sqrt{\br (Z_d \to \ell^+ \ell^-)} \label{BtoK1} \\
&&|\delta| \lsim 0.01 / \sqrt{\br (Z_d \to {\rm missing\; energy})} \label{BtoK2} .
\eea

It has been suggested \cite{Batell:2009jf} that even tighter bounds may be obtained from dedicated searches for $\ell^+ \ell^-$ pairs in $B$ decays, particularly if displaced vertices result from suppressed decay rates.
Nevertheless, even the relatively crude bounds in Eqs.~\eqref{BtoK1} and \eqref{BtoK2} are very constraining where applicable and are likely to be significantly improved by future dedicated searches.

On the basis of our analysis, it is clear that rare $K$ and $B$ decays provide sensitive windows to $Z$-$Z_d$ mass mixing and should be further explored in future high intensity experiments.
In fact for both cases, a more refined binned analysis of existing data would likely result in tighter bounds than those in Eqs.~\eqref{eq:n3a} and \eqref{BtoK1} or even uncover a hint of the $Z_d$'s presence.
Although applicable to a limited range of $m_{Z_d}$ and dependent on the $Z_d$ branching ratios, one can easily conclude $|\delta| \lsim 0.01 - 0.001$ over some restricted $m_{Z_d}$ domain.
In addition, further improvements are possible and warranted.
That constraint on $\delta$ sets a standard for other rare decay studies.
As we show in the next section, it is possible that searches for the rare Higgs decay $H \to Z Z_d$ have the statistical significance to also explore $|\delta| \lsim 0.01-0.001$ but have the potential advantage of covering a much broader range of $m_{Z_d}$ values including $m_{Z_d} \gsim 5 ~\gev$ if backgrounds can be controlled.

\section{Higgs Decays}
\label{sec:Hdecay}
\begin{table}[b]
\begin{tabular}{|c|c|}
\hline
$H$ Decay Channel                     & Branching Ratio \\
\hline\hline
$b \bar b$                            & $0.578$ \\
$W W^*$                               & $0.215$ \\
$gg$                                  & $0.086$ \\
$\tau^+ \tau^-$                       & $0.063$ \\
$c \bar c$                            & $0.029$ \\
$Z Z^*$                               & $0.026$ \\
$\gamma \gamma$                       & $2.3 \times 10^{-3}$ \\
$Z \gamma$                            & $1.5 \times 10^{-3}$ \\
\hline
$H \to Z Z^* \to \ell_1^+ \ell_1^- \ell_2^+ \ell_2^-$ & $1.2 \times 10^{-4}$ \\
$H \to Z Z^* \to \ell^+ \ell^- \nu \bar \nu$ ~~     & $3.6 \times 10^{-4}$ \\
\hline
\end{tabular}
\caption{Standard Model Higgs decay branching ratios for $m_H = 125 ~\gev$ ($\Gamma_H \simeq 4.1 ~\mev$) from Ref.~\cite{Hhandbook}.}
\label{tab:1}
\end{table}

We now address a primary consequence of our paper, the decay $H \to Z Z_d$ induced by $Z$-$Z_d$ mass matrix mixing.
To put our analysis into a current day perspective, we take $m_H = 125 ~\gev$, a value roughly suggested by early small excesses at the Large Hadron Collider (LHC) in the expected decay modes $H \to \gamma\gamma$, $WW^*$, $ZZ^*$ \cite{ATLAS:2012ae,Chatrchyan:2012tx}.
We note, however, that our findings regarding the sensitivity of Higgs searches for $H \to Z Z_d$ are fairly independent of the exact value of $m_H$.

To set the stage, we estimate that, roughly, one expects each LHC experiment to have about $75000$ Higgs bosons in the existing data before cuts (for the integrated luminosity of $4.7 - 4.9$ fb$^{-1}$ with $E_\text{c.m.} = 7 ~\tev$) for $m_H = 125 ~\gev$ in the SM.  
In Table \ref{tab:1}, we list the expected Higgs decay branching ratios within the context of the SM.
Of particular interest for comparison with $H \to Z Z_d$ are the SM decays (1) $H \to Z Z^* \to \ell_1^+ \ell_1^- \ell_2^+ \ell_2^-$ and (2) $H \to Z Z^* \to \ell^+ \ell^- \nu \bar \nu$ where the $*$ signifies a ``virtual,'' off mass shell boson and $\ell = e$, $\mu$.
The first of these, even at the $\br \sim 10^{-4}$ level, may have already been seen at the LHC where a handful of candidate events have been reported.
If it truly is a Higgs signal, hundreds more 4-lepton $\ell_1^+ \ell_1^- \ell_2^+ \ell_2^-$ events will be clearly observed in the coming years.  The second decay, $H\to \ell^+ \ell^- \nu \bar \nu$, is more difficult and to our knowledge has not been experimentally studied.   

\begin{figure}[t]
\begin{center}
\includegraphics[width=0.45\textwidth]{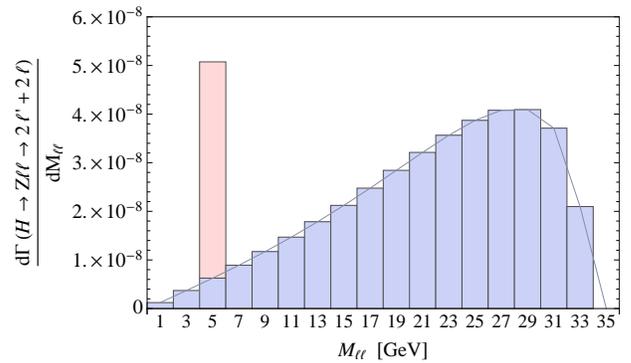}
\end{center}
\caption{Differential decay rate $H \to Z Z^* \to Z \ell^+ \ell^- \to 4 \ell$ vs $\ell^+ \ell^-$ invariant mass with $m_H = 125 ~\gev$ in the SM (in blue). For the illustration, $H \to Z Z_d \to Z \ell^+ \ell^-$ with $m_{Z_d} = 5 ~\gev$ and $\delta^2 \br (Z_d \to \ell^+ \ell^-) = 10^{-5}$ (which would need $N_\text{Higgs} \simeq 10^6$ for $3 \sigma$ evidence) is also shown (spike at the $5 ~\gev$ bin in red). Bin size is selected to be $2 ~\gev$.}
\label{fig:2}
\end{figure}

For the first case, one lepton pair will have an invariant mass of $m_Z \simeq 91 ~\gev$ while the second pair will have an invariant mass ranging from $0$ to about $34 ~\gev$ with a differential decay rate distribution as depicted in Fig.~\ref{fig:2}.
The second mode $H \to ZZ^* \to \ell^+\ell^-\nu\bar\nu$, with the neutrinos identified by missing energy, 
while experimentally more challenging should be searched for as well, since it can be used to constrain potentially invisible decays of the $Z_d$, as we subsequently discuss.

As we shall see, the decays $H \to Z Z_d$ are significantly enhanced beyond naive expectations, even for very small mixing.
To appreciate that phenomenon, we remind the reader that for a very heavy Higgs ($m_H^2 \gg m_W^2$, $m_Z^2$) the decay rates for $H \to W^+ W^-$ and $H \to Z Z$ can become enormous, growing like $\sim g^2 m_H^3 / m_V^2$, $V = W, Z$ with increasing $m_H$.
That behavior comes about because the final state $W$ and $Z$ bosons are longitudinally polarized, resulting in a $\sim m_H^2 / m_V^2$ enhancement factor at the decay rate level (for each final state gauge boson).

Such an effect is a manifestation of the Goldstone boson equivalence theorem which states that at high energies $(s \gg m_V^2)$, $S$-matrix elements involving $W^\pm$ and $Z$ bosons are equivalent, up to ${\cal O} (m_V / \sqrt{s})$, to the corresponding amplitudes in the Higgs-Goldstone scalar theory with the Goldstone boson replacing $W_L^\pm$, $Z_L$ (longitudinal components).
In the heavy Higgs limit, the $W^+ W^-$ and $Z Z$ decay products are essentially longitudinally polarized and behave like their Goldstone boson components.
The Higgs coupling to Goldstone bosons is of the form $(-ig/2) m_H^2 / m_V$, and squaring that coupling and dividing by $ 1/ m_H$ gives the $\Gamma (H \to VV) \sim g^2 m_H^3 / m_V^2$ exhibited by heavy Higgs decays.
We note that the longitudinal polarization of the gauge bosons can be very helpful in identifying a Higgs decay since the subsequent decay $W$ or $Z \to$ leptons have a characteristic angular distribution $\propto \sin^2\theta$ relative to the polarization.

Of course, our example of $125 ~\gev$ Higgs is too light to decay into $W^+ W^-$ or $ZZ$ pairs.
It can, however, decay into one real and one virtual boson with the latter directly producing a lepton pair with an invariant mass distribution as illustrated in Fig.~\ref{fig:2} \cite{Keung:1984hn}.
The integrated partial width for $H \to Z Z^* \to \ell_1^+ \ell_1^- \ell_2^+ \ell_2^-$ is, however, suppressed by $\alpha / 4\pi$ (from the $Z^* \ell_2^+ \ell_2^-$ coupling and 3-body phase space) and the small $\br (Z \to \ell^+ \ell^-) \simeq 2 \times 0.034$ for $\ell = e$, $\mu$.
One finds
\beq
\Gamma (H \to Z Z^* \to \ell_1^+ \ell_1^- \ell_2^+ \ell_2^-) \simeq 1.8 \times 10^{-6} \frac{G_F}{8 \sqrt{2} \pi} m_H m_W^2
\label{eq:19}
\eeq
with no significant sign of enhancement for longitudinal polarization, which is not surprising, since $m_H / m_Z \simeq 1.4$ in our example.
Nevertheless, even with the $10^{-6}$ suppression factor in Eq.~\eqref{eq:19}, it is expected that a SM $125 ~\gev$ Higgs should be starting to be seen with about several events per experiment in existing data, after acceptance cuts, 
and with hundreds more to follow in subsequent years.
So, Eq.~\eqref{eq:19} represents a decay rate standard that is easily discernible if backgrounds are in check.
We note that the decay rate for $H \to Z Z^* \to \ell^+ \ell^- \nu \bar\nu$ is expected in the SM to be about 3 times larger than Eq.~\eqref{eq:19} but more difficult to measure.

Now we come to the decay $H \to Z Z_d$ owing to $Z$-$Z_d$ mixing in our ``dark'' $Z$ scenario.
That mixing, parametrized by $\eps_Z = (m_{Z_d} / m_Z) \delta$, a very small quantity, might naively appear to be negligible since it leads to a tiny $H Z Z_d$ coupling $\sim (g / \cos\theta_W) m_Z \eps_Z$.
Consequently, the $H \to Z Z_d$ decay rate will be suppressed by $\eps_Z^2 = (m_{Z_d} / m_Z)^2 \delta^2$.
However, because of the Goldstone boson equivalence theorem, we gain an enhancement factor of $\sim (m_H / m_{Z_d})^2$ in the decay rate for longitudinally polarized $Z_d$ final states (a feature that may also help in identifying their subsequent $Z_d \to \ell_2^+ \ell_2^-$ products via angular distribution if statistics suffice).
That enhancement negates the small $m_{Z_d} / m_Z$ factor in the $H Z Z_d$ coupling.
Also, there is no $\alpha / 4\pi$ suppression for $H \to Z Z_d$, 
only the small $\br (Z \to \ell^+ \ell^-) \simeq 2 \times 0.034$ that needs to be included for $Z$ identification.
A detailed calculation (see Appendix B) leads to
\beq
\begin{split}
&\Gamma (H \to Z Z_d \to \ell_1^+ \ell_1^- \ell_2^+ \ell_2^-) \\
\simeq& 7 \times 10^{-3} \frac{G_F m_H^3}{8 \sqrt{2} \pi} \delta^2 \br (Z_d \to \ell_2^+ \ell_2^-)
\label{eq:20}
\end{split}
\eeq
Note the $m_H^3$ behavior that results from $Z$ and $Z_d$ being produced in their longitudinal polarization modes.
A similar formula with $\br (Z_d \to \ell_2^+ \ell_2^-)$ replaced by $\br (Z_d \to \text{missing energy})$ applies to the case $Z_d \to \nu \bar\nu$ or invisible ``dark'' particles.

In terms of its branching fraction relative to the SM expected width, one finds
\beq
\frac{\Gamma (H \to Z Z_d)}{\Gamma_H^\text{SM} (125 ~\gev)} \simeq 16 \times \delta^2 \lsim 0.1
\eeq
with $\Gamma_H^\text{SM} (125 ~\gev) \simeq 4.1 \times 10^{-3}$~GeV
\cite{Hhandbook} and using the low energy bound in Eq.~\eqref{eq:16}.
We see that as much as $10 \%$ of all LHC Higgs decays could be producing $Z Z_d$.
With current statistics, even a $10 \%$ loss of SM expectations 
would not be noticed; but eventually it would be uncovered by precision Higgs production and decay studies.

Taking the ratio of Eqs.~\eqref{eq:20} and \eqref{eq:19} gives 
\beq
\begin{split}
&\frac{\Gamma (H \to Z Z_d \to \ell_1^+ \ell_1^- \ell_2^+ \ell_2^-)}{\Gamma (H \to Z Z^* \to \ell_1^+ \ell_1^- \ell_2^+ \ell_2^-)} \\
\simeq& 10^4 \delta^2 \br (Z_d \to \ell_2^+ \ell_2^-)
\end{split}
\label{eq:21}
\eeq
with a similar expression
\beq
\begin{split}
&\frac{\Gamma (H \to Z Z_d \to \ell^+ \ell^- + \text{missing energy})}{\Gamma (H \to Z Z^* \to \ell^+ \ell^- + \text{missing energy})} \\
\simeq& (1/3) \times 10^4 \delta^2 \br (Z_d \to \text{missing energy})
\end{split}
\eeq
for invisible $Z_d$ decays.
Even for $\delta^2 \simeq 10^{-4}$, well below the atomic parity violation bound of $0.006$ in Eq.~\eqref{eq:16}, one would expect $H \to Z Z_d$ events with $\ell_1^+ \ell_1^- \ell_2^+ \ell_2^-$ or $\ell^+ \ell^- +$ missing energy to be starting to appear or already present in LHC data.
If there are no $Z_d \to$ dark particles decays, we expect the branching fractions of $Z_d$ into $\ell^+ \ell^-$ to be given by 
Eq.~\eqref{ZdBR}.
Therefore, in that case, one expects $\br (Z_d \to \ell^+ \ell^-)$ to be relatively large, particularly if $(\eps / \eps_Z)^2 \gsim 1$.
If $Z_d \to$ dark particles dominates its decay rate and significantly dilutes $\br (Z_d \to \ell^+ \ell^-)$, one still has the possibility of seeing $H \to Z Z_d \to \ell^+ \ell^- +$ missing energy, although this perhaps is more experimentally challenging.
Of course, given the original motivation for introducing a $Z_d$ into astrophysics as a way of explaining positron excesses through its decays, a relatively large $\br (Z_d \to \ell^+ \ell^-)$ might be expected.

Returning to Eq.~\eqref{eq:21}, we see that even for a somewhat suppressed $\br (Z_d \to \ell^+ \ell^-)$, the LHC experiments should be able to search for a $Z_d$ in the $H \to \ell_1^+ \ell_1^- \ell_2^+ \ell_2^-$ decay chain down to $\delta^2 \br (Z_d \to \ell^+ \ell^-) \sim {\cal O} (10^{-5})$, depending on backgrounds.
(The domain explored by rare $K$ and $B$ decays for some subset of $m_{Z_d}$ values.)
The signature, two isolated lepton pairs $\ell_1^+ \ell_1^- + \ell_2^+ \ell_2^-$ with a total invariant mass of $m_H$ and individual masses of $m_Z$ and $m_{Z_d}$ should stick out as a spike in the invariant mass plot of Fig.~\ref{fig:2}, as illustrated for $m_{Z_d} = 5 ~\gev$ and $\delta^2 \br (Z_d \to \ell^+ \ell^-) = 10^{-5}$.
In the bin centered at $M_{\ell\ell} = 5 ~\gev$, the SM expectation from Higgs of $m_H = 125 ~\gev$ is $\sim 6.3 \times 10^{-9} ~\gev$, while the signal associated with $H \to Z Z_d$ is $\sim 4.5 \times 10^{-8} ~\gev$.
With existing data of $N_H \simeq 75000$, no meaningful number of signal or background events are expected, and one would need $N_\text{Higgs} \simeq 10^6$ for $3 \sigma$ evidence (beyond the SM $H \to Z Z^* \to 4 \ell$ channel) at the LHC experiments.
However, this simple estimate ignores other reducible and irreducible backgrounds and a more reliable statement requires inclusion of such details.
Also, the $\ell_2^+ \ell_2^-$ decay pair from $Z_d$ should exhibit an angular distribution consistent with its longitudinal polarization.
That sensitivity is potentially orders of magnitude below the $\delta^2 < 0.006$ already established by atomic parity violation.  
We note that while the Higgs decay constraints on $\delta$ may not surpass those derived before
from rare $K$ and $B$ decays, they are applicable well beyond the $\ord{\gev}$ regime of $\mzd$, 
relevant for the meson decays. They represent a potentially unique broad capability of 
the LHC unmatched by low energy experiments.

We should point out that current searches for $H \to Z Z^* \to 4\ell$ are likely to miss $H \to Z Z_d$ because they generally cut out a lighter second lepton pair with 
$M_{\ell\ell} \lsim 15 ~\gev$, {\it i.e.} the range of interest, in order to avoid $Z \gamma^*$ backgrounds.
Hopefully, our results will provide some incentive for revisiting the low mass region in search of $Z_d$.

In addition to $Z_d \to \ell^+ \ell^-$, one should mount a search for $H \to Z Z_d \to \ell^+ \ell^- +$ missing energy.
Here, one might be helped by the fact that the missing energy and missing momentum of the $Z_d$ decay pair are nearly equal.
A thorough study of LHC capabilities for uncovering that decay mode is clearly warranted.  We also add that 
the Higgs can have a decay mode $H\to Z_d Z_d$, in our framework.  The rate for this decay is proportional to $\delta^4$, 
so, roughly, it is suppressed compared to the $Z Z_d$ mode by $\ord{\delta^2}$ which, 
given our bound in Eq.~(\ref{eq:16}), is a suppression of $0.006$ or smaller.  
The rate for the $Z_d Z_d$ channel could be enhanced if hidden sector 
scalars that couple directly to $Z_d$ and give it mass are 
allowed to mix with the SM sector Higgs scalars.

\begin{table*}[t]
\begin{tabular}{|c|c|c|}
\hline
Process & Current (future) bound on $\delta$ & Comment \\
\hline\hline
Low Energy Parity Violation & $|\delta| \lsim 0.08 - 0.01 ~(0.001)$ & Fairly independent of $m_{Z_d}$. Depends on $\eps$.  \\
Rare $K$ Decays & $|\delta| \lsim 0.01 - 0.001 ~(0.0003)$      & $m_\pi^2 < m_{Z_d}^2 \ll m_K^2$. Depends on $\br(Z_d)$. \\
Rare $B$ Decays & $|\delta| \lsim 0.02 - 0.001 ~(0.0003)$      & $m_\pi^2 < m_{Z_d}^2 \ll m_B^2$. Depends on $\br(Z_d)$. Some mass gap $\sim 3 ~\gev$. \\
$H \to Z Z_d$   & $|\delta| \lsim (0.003 - 0.001)$             & $m_{Z_d}^2 \ll (m_H - m_Z)^2$. Depends on $\br(Z_d)$ and background. \\
\hline
\end{tabular}
\caption{Rough ranges of current (future) constraints on $\delta$ from various processes examined along with commentary on applicability of the bounds.
These processes have negligible sensitivity to pure kinetic mixing effects.}
\label{tab:2}
\end{table*}

\section{A 2 Higgs Doublet Example}
\label{2HD}
In the preceding discussion, we examined the dark $Z$ phenomenology in a general framework.  As mentioned before, 
the main ingredient we introduced was mass mixing between the SM $Z$ and the $Z_d$ which could be realized in a variety 
of models.  In this section, to demonstrate how our general framework might be realized, we will consider a 2 Higgs doublet extension of the SM. 
(See Ref.~\cite{Branco:2011iw} for a recent review on 2HD models.)
Here, we assume two $SU(2)_L \times U(1)_Y$ 
Higgs doublets, $H_1$ and $H_2$, but allow $H_2$ to carry a ``dark'' charge that couples it directly to $U(1)_d$.  Note that the assumption of the $U(1)_d$ in our example is well-motivated, as it allows 
the model to evade severe constraints from flavor-changing neutral currents that are often addressed through the 
introduction of a $\mathbb{Z}_2$ symmetry in generic 2HD models.  We also allow, for generality, 
a singlet scalar, $H_d$, that also provides part of the $Z_d$ mass through its 
``dark'' sector vacuum expectation value $v_d$. 

With the above assumptions, $H_2$ does not couple directly to ordinary 
fermions, but does contribute to $W^\pm$, $Z$ and $Z_d$ 
masses as well as $Z$-$Z_d$ mixing through its vacuum expectation value $v_2$.  
Such a setup is akin to what is often called a Type-I 2HD model~\cite{Hhunter}. Here, we will 
take $H_1$ to be a SM-like Higgs scalar, identified as $H$ in our preceding general analysis.  
To keep the discussion simple, we ignore scalar mixing among the $H_1$, $H_2$, and $H_d$ states.  
The $v_1$, $v_2$, and $v_d$ vacuum expectation values of $H_1$ (the SM doublet), $H_2$ and $H_d$ give rise to $\delta = \sin\beta \sin\beta_d$ where $\tan\beta = v_2 / v_1$ and $\tan\beta_d = v_2 / v_d$, as will be shown in Appendix B.  The condition  
of a SM-like $H_1$ can be satisfied, to a good approximation, for $\tan\beta \lsim 1/3$, and does not require a large hierarchy 
of scales in the Higgs sector.   
The constraints on $\delta$ previously discussed will however constrain the product $\sbsd$.

There are many additional features of our 2HD model worth studying.
For example, nonzero Higgs scalar mixing (which we set to zero) could give 
rise to enhancements in $H \to Z_d Z_d$, as mentioned before, or perhaps $H \to h h$ 
($h$ being a lighter Higgs scalar remnant of $H_2$) \cite{Ferreira:2012my}.
Those possibilities are interesting but more model dependent.

\section{Summary and Conclusion}
In this work, we explored the possibility of mass mixing between the $Z$ boson of the SM and 
a new light vector boson $Z_d$ associated with a hidden or dark sector $U(1)_d$ gauge symmetry.  Such a light state 
has been invoked in discussions of astrophysical anomalies that may originate from cosmic dark matter.  We dub this 
new vector boson the ``dark'' $Z$, as its properties are analogous to that of the SM $Z$.  In particular, the couplings 
of $Z_d$ can provide new sources of parity violation and measurably affect the decay of the Higgs through 
novel channels such as $H \to Z Z_d$.  Existing atomic parity violation, polarized $e$ scattering, and 
rare $K$ and $B$ decay data 
already place interesting bounds on the degree of $Z$-$Z_d$ mass mixing, but further improvement 
is possible and warranted (see Table~\ref{tab:2})\footnote{
One could contemplate searching for $Z_d$ effects in precision neutrino neutral current cross section measurements such as $\nu_\mu e \to \nu_\mu e$ or deep-inelastic $\nu_\mu N \to \nu_\mu X$.
However, to be competitive with anticipated low energy parity violation polarized electron scattering or atomic experiments, those neutrino studies would have to reach $\sim \pm 0.1\%$ statistical and normalization uncertainties, a challenging task that would likely require a high energy neutrino factory (see Ref.~\cite{Marciano:2003eq}).
A detailed discussion of $Z_d$ effects on neutrino cross sections will be given in a separate publication.
}.

The presence of kinetic 
mixing affects the phenomenology of $Z_d$, but much of the main physics discussed in our work 
persists even in the absence of kinetic mixing.  Various experimental efforts are currently devoted to possible signals 
of the ``dark" photon, based solely on the possibility of kinetic mixing between $U(1)_d$ and the SM photon.  Here, we want to emphasize the $m_Z^2 / m_{Z_d}^2$ enhancement factor in low energy parity violation and the longitudinal polarization enhancement $E_{Z_d} / m_{Z_d}$, with $E_{Z_d}$ the energy of $Z_d$, in rare meson decays and the Higgs decay $H \to Z Z_d$.
These enhancements make such processes particularly sensitive to very small $Z$-$Z_d$ mixing.  In particular, 
future polarized $ep$ and $ee$ scattering experiments can provide further probes of the scenario we have considered 
in this work.  These parity violating probes are sensitive to a wide range of $Z_d$ masses, including $m_{Z_d} \lsim 140 ~\mev$, where other searches fail because of $\pi^0$ Dalitz decays background and are independent of $Z_d$ branching fractions.  The rare $K$ and $B$ decays currently provide some of the most 
stringent bounds on the degree of $Z$-$Z_d$ mixing, however they depend on the $Z_d$ branching fractions 
and also do not apply to $\mzd$ above the meson mass.
In addition, there can be gaps in the bounds, for example in the $m_{Z_d}$ charmonium mass region.

In the event of the discovery of a SM-like Higgs at the LHC, say at $\sim 125~\gev$ based on current hints, a new front in the search for a dark $Z$ can be established.
The Higgs decay data are particularly unique for $\mzd \gsim 5~\gev$, and hence probe a part of parameter 
space that is inaccessible to meson data.  The reach for this new physics can be extended well beyond the 
current limits through precise measurements of Higgs decays, as may be done at an $e^+ e^-$ or $\mu^+ \mu^-$ collider if high statistics are available.  We conclude that pushing the above types of experiments as far as possible is strongly motivated, for they 
could be windows to the ``dark side'' of particle physics.

\vskip3mm
\acknowledgments
This work was supported in part by the United States Department of Energy under Grant No. DE-AC02-98CH10886.
WM acknowledges partial support from the Gutenberg Research College.

\newpage
\appendix
\begin{widetext}

\section{Gauge Kinetic Terms}
The gauge kinetic terms allowed by the gauge symmetries $SU(2)_L \times U(1)_Y \times U(1)_d$ are
\beq
{\cal L}_\text{gauge} = - \frac{1}{4} \hat B_{\mu\nu} \hat B^{\mu\nu} + \frac{1}{2} \frac{\eps}{\cos\theta_W} \hat B_{\mu\nu} \hat Z_d^{0 \mu\nu} - \frac{1}{4} \hat Z_{d \mu\nu}^0 \hat Z_d^{0 \mu\nu}
\eeq
with $F_{\mu\nu} = \partial_\mu F_\nu - \partial_\nu F_\mu$.
The hatted quantities are fields before the diagonalization of the gauge kinetic terms.
The diagonalization is done by the field redefinition known as a $GL(2,R)$ rotation
\beq
\mat{Z_{d \mu}^0 \\ B_\mu} = \mat{\sqrt{1-\eps^2/\cos^2\theta_W} && 0 \\ -\eps/\cos\theta_W && 1} \mat{\hat Z_{d \mu}^0 \\ \hat B_\mu}
\label{eq:A2}
\eeq
after which, $B$ gets a $\hat Z_d$ component proportional to $\eps$ while $Z_d$ does not get any $\hat B$ component.
\beq
{\cal L}_\text{gauge} = - \frac{1}{4} B_{\mu\nu} B^{\mu\nu} - \frac{1}{4} Z_{d \mu\nu}^0 Z_d^{0 \mu\nu}
\eeq

We will take $\hat Z_{d \mu}^0 = Z_{d \mu}^0$ and $\hat B_\mu = B_\mu + (\eps / \cos\theta_W) Z_{d \mu}^0$ and ignore ${\cal O}(\eps^2)$ terms from here on.
After electroweak mixing with Weinberg angle $\theta_W$
\beq
\mat{A \\ Z^0} = \mat{\cos\theta_W && \sin\theta_W \\ -\sin\theta_W && \cos\theta_W} \mat{B \\ W_3}
\label{eq:A4}
\eeq
we get
\beq
\begin{split}
A_\mu       &= \hat A_\mu - \eps \hat Z_{d \mu}^0 \\
Z_\mu^0     &= \hat Z_\mu^0 + \eps \tan\theta_W \hat Z_{d \mu}^0 \\
Z_{d \mu}^0 &= \hat Z_{d \mu}^0
\end{split}
\label{eq:A5}
\eeq
as an effect of the gauge kinetic mixing.
Thus, $Z_d^0$ is unaffected to $\ord{\eps}$ while both $A_\mu$ and $Z_\mu^0$ are shifted by the gauge kinetic mixing followed by the electroweak mixing.
However, the bare fields do not take into consideration $Z^0$-$Z^0_d$ mixing via the mass matrix from the Higgs mechanism which we will deal with in the following.

\section{Scalar Kinetic Terms}
The scalar kinetic term is given by
\beq
{\cal L}_\text{scalar} = \sum_i | D_\mu \Phi_i |^2
\label{eq:Lke}
\eeq
where $i$ runs for all Higgs scalars.
Considering only neutral components of gauge bosons, we have
\beq
D_\mu \Phi_i = \left( \partial_\mu + i g' Y[\Phi_i] \hat B_\mu + i g T_3[\Phi_i] \hat W_{3 \mu} + i g_d Q_d[\Phi_i] \hat Z_{d \mu}^0 \right) \Phi_i
\eeq
before gauge kinetic diagonalization where $Y$, $T_3$, and $Q_d$ are hypercharge, isospin, and dark charge, respectively.

After symmetry breaking, the scalars can be written with the vacuum expectation values ($v_i$).
\beq
\Phi_i = \frac{1}{\sqrt{2}} \left( H_i + v_i \right)
\eeq

\subsection{Vector boson mass}
From Eq.~\eqref{eq:Lke}, we can get the relevant vector boson mass terms
\beq
{\cal L}_\text{scalar} = \frac{1}{2} m_{Z^0}^2 Z^0 Z^0 - \Delta^2 Z^0 Z_d^0 + \frac{1}{2} m_{Z_d^0}^2 Z_d^0 Z_d^0 + \cdots .
\eeq
The mixing of two vector bosons is given by
\beq
\mat{Z \\ Z_d} = \mat{ \cos\xi && -\sin\xi \\ \sin\xi && \cos\xi } \mat{Z^0 \\ Z_d^0}
\eeq
with
\beq
\begin{split}
\tan 2\xi &= \frac{2 \Delta^2}{m_{Z^0}^2 - m_{Z_d^0}^2} .
\end{split}
\eeq

\vskip3mm
\noindent
{\it 2HD Model Realization:} \\

\noindent
We discuss some details in context of the 2HD model example in Sec.~\ref{2HD}.
We set $U(1)_d$ charges as $Q_d[H_1] = 0$, $Q_d[H_2] = Q_d[H_d] = 1$ for notational convenience.
Then the gauge boson mass-squared is given by, with $g_Z = g' / \sin\theta_W = g / \cos\theta_W$,
\beq
\begin{split}
m_{Z^0}^2  &= \frac{1}{4} g_Z^2 (v_1^2 + v_2^2) , \\
m_{Z_d^0}^2 &= g_d^2 (v_2^2 + v_d^2) + \frac{\eps}{\cos\theta_W} g_d g' v_2^2 + \frac{1}{4} \left( \frac{\eps}{\cos\theta_W} \right)^2 g'^2 (v_1^2 + v_2^2) , \\
\Delta^2   &= \frac{1}{2} g_d g_Z v_2^2 + \frac{1}{4} \frac{\eps}{\cos\theta_W} g_Z g' (v_1^2 + v_2^2) .
\end{split}
\label{eq:B7}
\eeq

We assume $m_{Z_d^0}^2 \ll m_{Z^0}^2$ which will be the case as long as ($g_d^2$, $\eps g_d$, $\eps^2$) $\ll g_Z^2$ and $v_d$ is not exceedingly larger than the electroweak scale.
We define $\tan\beta \equiv v_2 / v_1$, $\tan\beta_d \equiv v_2 / v_d$, and $v^2 \equiv v_1^2 + v_2^2 \simeq (246 ~\gev)^2$.
Then we have
\beq
\begin{split}
m_Z^2     &\simeq m_{Z^0}^2 = \frac{1}{4} g_Z^2 v^2 , \\
m_{Z_d}^2 &\simeq m_{Z_d^0}^2 - \frac{(\Delta^2)^2}{m_{Z^0}^2} = g_d^2 (v_d^2 + v^2 \sin^2\beta \cos^2\beta) = g_d^2 v^2 \frac{\sin^2\beta}{\sin^2\beta_d} (1 - \sbsdSqr) , \\
\xi       &\simeq \frac{\Delta^2}{m_{Z^0}^2} = \frac{2 g_d}{g_Z} \sin^2\beta + \eps \tan\theta_W .
\end{split}
\label{eq:B9}
\eeq
Gauge kinetic mixing $\eps$ does not contribute to $Z_d$ mass but it affects the $Z$-$Z_d$ mixing angle $\xi$.

\vskip3mm
(i) In the $v_2 = 0$ limit ({\it i.e.} pure dark photon limit), the $Z_d$ mass is entirely from the Higgs singlet $H_d$ and the $Z$-$Z_d$ mixing angle is provided entirely by $\eps$.
We have
\beq
m_Z^2 \simeq m_{Z^0}^2 \qquad m_{Z_d}^2 \simeq g_d^2 v_d^2 \qquad \xi \simeq \eps \tan\theta_W
\eeq
which give
\beq
M_0^2 \simeq \mat{m_Z^2 & -\eps \tan\theta_W m_Z^2 \\ -\eps \tan\theta_W m_Z^2 & m_{Z_d}^2 + \eps^2 \tan^2 \theta_W m_Z^2 } .
\eeq
The mixing induced by the mass matrix cancels the effects because of field redefinition in Eq.~\eqref{eq:A5} for the $Z_d$ induced neutral current coupling.

\vskip3mm
(ii) In the $\eps = 0$ limit ({\it i.e.} pure dark $Z$ limit),
\beq
\begin{split}
m_{Z_d}^2 &\simeq g_d^2 v^2 \frac{\sin^2\beta}{\sin^2\beta_d} (1 - \sbsdSqr) \\
\xi       &\simeq \frac{2 g_d}{g_Z} \sin^2\beta \simeq \frac{m_{Z_d}}{m_Z} \frac{\sbsd}{\sqrt{1 - \sbsdSqr}} .
\end{split}
\label{eq:newB11}
\eeq
Taking $1 - \sbsdSqr \simeq 1$ is valid when $|\Delta^2| \ll m_{Z^0} m_{Z_d^0}$.
In this limit
\beq
m_Z^2 \simeq m_{Z^0}^2 \qquad m_{Z_d}^2 \simeq m_{Z_d^0}^2 \qquad \xi \simeq \eps_Z
\label{eq:newB12}
\eeq
with
\beq
\eps_Z = \frac{m_{Z_d}}{m_Z} \delta ~~~\text{and}~~~ \delta = \sbsd.
\eeq

\subsection{Higgs-Vector-Vector Couplings}
We assume no mixing among Higgs scalars and refer to the SM-like Higgs as $H$.
From Eq.~\eqref{eq:Lke}, we can get the relevant Higgs coupling to vector bosons.
\beq
{\cal L}_\text{scalar} = \frac{1}{2} {\cal C}_{H Z Z} H Z Z + {\cal C}_{H Z Z_d} H Z Z_d + \frac{1}{2} {\cal C}_{H Z_d Z_d} H Z_d Z_d + \cdots
\eeq
The Feynman rules for coupling of $H$ to two vector bosons $V_1$ and $V_2$ are then given by $i g_{\mu\nu} {\cal C}_{H V_1 V_2}$.

\vskip3mm

In the 2HD example, we get
\beq
\begin{split}
{\cal C}_{H Z Z}     &= {\cal C}_{H Z Z}^\text{SM} \cos\beta (\cos\xi + \eps \tan\theta_W \sin\xi)^2 \\
{\cal C}_{H Z Z_d}   &= {\cal C}_{H Z Z}^\text{SM} \cos\beta (\cos\xi + \eps \tan\theta_W \sin\xi) (\sin\xi - \eps \tan\theta_W \cos\xi) \\
{\cal C}_{H Z_d Z_d} &= {\cal C}_{H Z Z}^\text{SM} \cos\beta (\sin\xi - \eps \tan\theta_W \cos\xi)^2
\end{split}
\label{eq:B18}
\eeq
with ${\cal C}_{H Z Z}^\text{SM} = \frac{1}{2} g_Z^2 v$.

The ratio of couplings is
\beq
\Theta = \frac{{\cal C}_{H Z Z_d}}{{\cal C}_{H Z Z}} = \frac{{\cal C}_{H Z_d Z_d}}{{\cal C}_{H Z Z_d}} = \frac{\sin\xi - \eps \tan\theta_W \cos\xi}{\cos\xi + \eps \tan\theta_W \sin\xi} .
\eeq
which, with small $|\xi| \ll 1$ from Eq.~\eqref{eq:B9}, yields
\beq
\Theta \simeq \xi - \eps \tan\theta_W \simeq \frac{2 g_d}{g_Z} \sin^2\beta
\eeq
showing that $\Theta$ is not sensitive to $\eps$.

The relevant Higgs decay rates, for $m_{Z_d} \ll m_H$, are given by
\beq
\begin{split}
\Gamma(H \to Z Z)     &= \frac{1}{128 \pi} \frac{m_H^3}{m_Z^4} \sqrt{1 - \frac{4 m_Z^2}{m_H^2}} \left( 1 - \frac{4 m_Z^2}{m_H^2} + \frac{12 m_Z^4}{m_H^4} \right) \left({\cal C}_{HZZ} \right)^2 \\
\Gamma(H \to Z Z_d)   &\simeq \frac{1}{64 \pi} \frac{m_H^3}{m_Z^2 m_{Z_d}^2} \left( 1 - \frac{m_Z^2}{m_H^2} \right)^3 \left(\Theta {\cal C}_{HZZ} \right)^2 \\
\Gamma(H \to Z_d Z_d) &\simeq \frac{1}{128 \pi} \frac{m_H^3}{m_{Z_d}^4} \left(\Theta^2 {\cal C}_{HZZ}\right)^2
\end{split}
\label{eq:D2}
\eeq
with couplings given in Eq.~\eqref{eq:B18}.
Eq.~\eqref{eq:D2} conveniently shows the effects of phase space and $Z$-$Z_d$ mixing in the Higgs decay rates.
The ratio of Higgs decay rates in the $Z_d Z_d$ and $Z Z_d$ channels is
\beq
\begin{split}
\frac{\Gamma(H \to Z_d Z_d)}{\Gamma(H \to Z Z_d)} &\simeq \frac{\Theta^2}{2} \frac{m_Z^2}{m_{Z_d}^2} \left( 1 - \frac{m_Z^2}{m_H^2} \right)^{-3} \\
&\simeq \frac{1}{2} \sbsdSqr \left( 1 - \frac{m_Z^2}{m_H^2} \right)^{-3} = \frac{1}{2} \delta^2 \left( 1 - \frac{m_Z^2}{m_H^2} \right)^{-3}
\end{split}
\label{eq:D3}
\eeq
where Eqs.~\eqref{eq:newB11} and ~\eqref{eq:newB12} have been used in the second line.

\section{$Z_d$ Decay Width}
Using Eqs.~\eqref{eq:8} and \eqref{eq:9} in the text, we find that the partial decay width of $Z_d$ into the SM fermion pair $f \bar f$ is given by, neglecting $m_f / m_{Z_d}$ corrections \cite{Albert:1979ix},
\beq
\Gamma (Z_d \to f \bar f) \simeq \frac{N_C}{48 \pi} \eps_Z^2 g_Z^2 \left( g'^2_{V f} + g^2_{A f} \right) m_{Z_d} ,
\eeq
where $g'_{V f} = T_{3 f} - 2 Q_f \left(\sin^2\theta_W - (\eps / \eps_Z) \cos\theta_W \sin\theta_W \right)$ and $g_{A f} = -T_{3 f}$.
Here, $N_C = 3$ for quarks and $N_C = 1$ for leptons.

\end{widetext}



\end{document}